# Sub-cycle control of terahertz waveform polarization using all-optically induced transient metamaterials

N. Kamaraju[1,2], Andrea Rubano[1,3], Linke Jian[4], Surajit Saha[4], T. Venkatesan[4], Jan Nötzold[1], R. Kramer Campen[1], Martin Wolf[1], Tobias Kampfrath[1]

[1] Fritz Haber Institute of the Max Planck Society, Faradayweg 4-6, 14195 Berlin, Germany.
[2] Department of Electrical, Computer, and Systems Engineering, Rensselaer Polytechnic Institute, Troy, New York 12180, USA.
[3] CNR-SPIN, Dipartimento di Fisica, Università di Napoli Federico II, Monte S. Angelo, Via Cintia, Napoli, Italy.
[4] NUSNNI-Nanocore, National University of Singapore, Singapore.

**Coherent radiation with frequencies ranging from 0.3 to 30 THz has recently become accessible by femtosecond laser technology. Terahertz (THz) waves have already found many applications in spectroscopy and imaging, and they can be manipulated using static optical elements such as lenses, polarizers, and filters. However, ultrafast modulation of THz radiation is required as well, for instance in short-range wireless communication or for preparing shaped THz transients for coherent control of numerous material excitations. Here, we demonstrate an all-optically created transient metamaterial that permits to manipulate the polarization of THz waveforms with sub-cycle precision. The polarization-modulated pulses are potentially interesting for controlling elementary motions such as vibrations of crystal lattices, rotations of molecules, and the precession of spins.**

Coherent electromagnetic transients covering the terahertz (THz) frequency window can nowadays routinely be generated and detected using femtosecond laser technology[1]. Such THz pulses have numerous applications, an example of which is the probing[2] and even control[3] of fundamental modes like vibrations of molecules and crystals as wells as spin precessions in solids. In addition, THz radiation has proven as an excellent tool in non-destructive and material-specific imaging[4] and is under consideration as an information carrier in short-range wireless communication at THz bit rates, potentially thereby replacing slower current gigahertz-based technology[5]. Such applications heavily rely on the manipulation of THz radiation[6], and static elements like lenses, waveplates and frequency filters are readily available[7]. However, dynamical manipulation of THz transients with sub-cycle precision is required as well in fields like THz telecommunication (where information has to be imprinted on a pulse train[5]) and THz coherent control over matter (where tailored waveforms have been predicted to drive, for instance, the configuration of a crystal lattice along a desired path[8]).

To date, ultrafast THz switches and modulators have been demonstrated based on microstructured photonic structures with enhanced light-matter interaction[9,10,11,12,13,14,15,16,17]. Amongst those, metamaterials, artificial plane metallic structures on an insulating substrate with a period much smaller than the vacuum wavelength of the THz radiation (300μm at 1THz), have an effective refractive index that can be tuned by tailoring the shape of the metal structure. Ultrafast switching is triggered by photodoping the underlying semiconductor substrate through a femtosecond laser pulse, thereby modifying its refractive index. However, apart from such global modulation, the geometry of the metal structure is difficult to change, and the few constituting materials do not allow for a gradual spatial variation of the refractive index. A more flexible approach is to start with a homogeneous substrate into which the photonic structure is imprinted all-optically[18,19,20,21], for example with the scheme shown in Figure 1a. Here, a laser beam with a spatially modulated cross section is incident onto a plane substrate in which the optical intensity pattern is translated into a spatial modulation of the refractive index. Compared to standard lithography, such "gray-scale transient lithography" offers high flexibility. First, when combined with a freely programmable spatial light modulator[19], the transverse modulation of the pump beam can be changed completely. Second, one can realize a continuous variation of the optical permittivity, on ultrafast time scales. So far, this scheme has been used to imprint transient gratings with a period of the order of the probing wavelength which permitted the partial diffraction of an incident infrared[18] or THz pulse[19,20] into a surface-type or free-space wave. However, fascinating prospects such as transient metamaterials and their feasibility for on-the-fly manipulation of the THz transient field have not yet been implemented.

In this report, we demonstrate the ultrafast all-optical generation of a transient wire-grid polarizer (WGP), a textbook example of an anisotropic metamaterial, and its use for the manipulation of the polarization state of THz transients with sub-cycle precision. Potential applications of this approach are ultrafast pulse splitters and polarization multiplexers which are, for example, required in THz wireless communication technology. Moreover, the scheme can straightforwardly be extended to implement all varieties of transient THz photonic devices like lenses, filters, and wave plates.

To this end, we make use of the scheme shown in Figure 1a and irradiate a semiconductor slab with a visible femtosecond laser pulse (center wavelength 800nm) whose transverse beam profile can be shaped arbitrarily by a spatial light modulator. Through light absorption in the semiconductor, the beam pattern is translated into a spatial density distribution of an electron-hole plasma and, thus, refractive-index distribution $n(x,y,z,\omega)$ in the slab: irradiated points $(x,y,z)$ become metallic (real part of the permittivity $n^2$ is negative) whereas dark areas remain dielectric (Re $n^2>0$). The effect of the photoinduced charge carriers is more pronounced at lower frequencies $\omega/2\pi$ because their permittivity is roughly proportional to the pump intensity times $1/\omega^2$ (see Supplementary Figure S1). Since features much smaller than the THz wavelength can be realized with optical radiation, we can "switch on" a tailored metamaterial for THz radiation on the femtosecond time scale. The transmittance of this dynamic structure is probed by a THz pulse arriving at a variable delay after the pump pulse. By temporally overlapping the pump and the probe beam, we are even able to manipulate the THz transient with sub-cycle precision (Figure 1a).

As a fundamental test structure, we choose the textbook example of a WGP composed of a periodic array of parallel metal wires. This pattern is realized by imaging a shadow mask with a periodic arrangement of adjacent fully transparent and absorbing stripes onto an insulating Si slab (Figure 1a). The measured (Figures 1b and 1c) and calculated (Figure 1c) beam profile in the slab plane shows that the distance of adjacent stripes (30µm) is substantially smaller than the THz wavelength inside Si (88µm at 1THz). Thus, we expect that our transient WGP acts as an effective homogeneous anisotropic medium whose optical properties are fully determined by the refractive index along its symmetry axes. Note that the pump intensity pattern (red symbols in Figure 1c) is not as perfectly rectangular as the shadow mask, which is a result of the 12µm wide point spread function of the telescope imaging the mask onto the Si surface. The slab thickness of 10µm approximately equals the penetration depth of the pump radiation (wavelength 800nm) such that the excitation profile is virtually homogeneous along z, that is, $n=n(x,y,\omega)$. In Si, the pump-induced change in the refractive index proceeds in a temporally step-like manner, with a rise time of ~100fs and a relaxation on a nanosecond time scale[22].

When the THz probe pulse is polarized parallel (∥) to the wires, free charge carriers inside the metallic parts are moved in the same way as for a plane metal surface. Likewise, the substrate parts between wires respond in the same way as an extended dielectric slab. This configuration can be considered as a parallel connection of electric resistors, and the effective refractive index $n_\parallel$ for this configuration is given by the relation[23]

$$n_\parallel^2(\omega) = \langle n^2(x,y,\omega)\rangle \qquad (1)$$

where $\langle . \rangle$ denotes averaging along the in-plane direction perpendicular to the wires. The situation is very different when the incident THz radiation is polarized perpendicular (⊥) to the wires. In this case, charge carriers will be shifted towards the wire surface which leads to the buildup of an electric field that eventually compensates the driving field inside the wire. As a consequence, much less charge than for parallel polarization will be translated, implying less reflection of the incident light. This configuration can be considered as a series connection of resistors, and the effective refractive index $n_\perp$ of the metamaterial is determined by[23]

$$\frac{1}{n_\perp^2(\omega)} = \left\langle \frac{1}{n^2(x,y,\omega)} \right\rangle. \qquad (2)$$

The electromagnetic properties of the WGP are determined by broadband THz probe pulses polarized either along the ∥ or the ⊥ direction (Figure 1a). As a reference, Figure 2a shows the transient electric field $E_{UE}(t)$ of a THz pulse detected after having traversed an unexcited (UE) silicon slab. When the slab is excited with a femtosecond laser pulse with a homogeneous beam profile (Figure 1c) approximately 60ps before arrival of the probe pulse, we observe that the THz field amplitude undergoes a drastic reduction by about one order of magnitude (Figure 2b). This decrease predominantly arises from the large reflectance (rather than the absorption) of the transiently metallized Si slab, which is found to increase from 53% for an unexcited slab to 86% at 1THz (see Supplementary Figure S2).

It is interesting to observe what happens when the homogeneous excitation profile is replaced by the striped structure of Figure 1b. When the stripes are parallel (∥) to the THz polarization, we obtain a waveform (Figure 2c) whose overall amplitude is a factor of ~2 larger than that observed for homogeneous excitation (Figure 2b). This rescaling effect is plausible since the effectively excited Si volume fraction has been reduced by approximately the same factor (Figure 1c). Note, however, the strikingly different behavior when the stripe direction is chosen perpendicular (⊥) to the THz polarization: in this case, the field amplitude is a factor of ~3 larger than for the ∥ configuration (Figure 2d). In other words, the transient structure indeed acts like a polarizer with a power extinction ratio of the order of 10.

The highly anisotropic response of the transient metamaterial can be studied further in the frequency domain: Figure 3a displays the power transmittance $|E_\perp(\omega)/E_{UE}(\omega)|^2$ and $|E_\parallel(\omega)/E_{UE}(\omega)|^2$, that is, the intensity spectrum of the respective perpendicular and parallel

component normalized to the spectrum of the unexcited slab. While the ∥ transmittance is found to increase monotonously with frequency, its ⊥ counterpart exhibits a broad maximum. In addition, we observe that the transmittance of the transient WGP is significantly larger for perpendicular than for parallel polarization, reaching an extinction ratio $|E_\perp(\omega)/E_\parallel(\omega)|^2$ of more than 10 around $\omega/2\pi=1.5$THz (Figure 3a), in agreement with our time-domain observations (Figure 2). Interestingly, the ∥ component is found to exhibit a larger negative, more metal-like phase shift than the ⊥ component (Figure 3b). Therefore, our transient WGP can also be used as a retardation plate.

To show that the photoexcited slab can indeed be considered as a birefringent THz metamaterial, we calculate the refractive indices $n_\perp(\omega)$ and $n_\parallel(\omega)$ using effective-medium theory[23] [equations (1) and (2)]. The refractive-index distribution $n(x,y,z,\omega)$ is obtained from Figure 2b and measurements of the Si refractive index after homogeneous excitation with various pump fluences (see Supplementary Figure S1). Using $n_\perp(\omega)$ and $n_\parallel(\omega)$, the transmission coefficients of the slab are inferred using standard formulas[24] that also account for all reflections occurring at the Si-air interfaces. Modulus and phase of the calculated transmission coefficient are shown by the solid lines in Figures 3a and 3b, respectively. We find excellent agreement between experiment and theory, both in terms of magnitude and trend, without invoking any fit parameters. It is interesting to note that this agreement extends to frequencies of 2.5THz at which the metamaterial condition (periodicity much smaller than wavelength inside medium) is no longer entirely met. We summarize that our modeling corroborates the notion that we have successfully and all-optically created a transient metamaterial.

While our experiment provides a first proof of principle of a photoinduced metamaterial, our model allows us to predict the performance of the WGP under modified conditions. More precisely, calculations indicate that a substantially enhanced extinction ratio (up to 4000 at 0.6 THz), enhanced relative ∥-transmittance (more than 100%), and increased bandwidth could be achieved by cooling the Si slab to cryogenic temperatures (thereby reducing THz losses arising from Drude absorption[25]) and by increasing the contrast between dark and bright regions of the pump intensity pattern (see Supplementary Figure S3).

As our metamaterial is generated on a time scale of ~100fs, it offers the fascinating possibility for on-the-fly polarization manipulation of THz transients, with sub-cycle precision. The results of such an experiment is shown in Figure 4. A THz pulse, polarized along $y$ and, thus, 45° with respect to the WGP major axes, is incident onto the still unexcited Si slab (see Figure 1a). Once part of the transient has traversed the Si, the pump pulse imprints the WGP structure, as a consequence of which the ∥ and ⊥ components of the remaining THz transient experience a different attenuation and phase shift, resulting in an elliptically polarized trailing part of the THz pulse.

In conclusion, we have provided a first experimental demonstration of an ultrafast all-optical metamaterial that permits the manipulation of THz transients with sub-cycle resolution. The polarization-modulated pulses are potentially interesting for controlling elementary motions such as lattice vibrations[26,27], molecular rotations[28], and spin precession[29]. Future work will employ programmable spatial light modulators[19] in order to realize more complex transient metamaterials that will enable on-the-fly manipulation of THz light such as frequency shifting[30] and spectral lensing[31].

# Methods

We employ intense femtosecond laser pulses (duration 40fs, center wavelength 800nm, pulse energy 7mJ, repetition rate 1kHz), part of which is used to generate THz pulses by optical rectification in ZnTe(110) crystals (thickness 0.5mm). The major part of the laser beam (diameter 1cm) is sent through a glass substrate on top of which an array of Cr wires (thickness 103nm, width 20µm, period length 30µm) is fabricated by means of optical lithography. This pattern is imaged onto a high-resistivity Si slab (thickness 10µm) through a 4$f$-imaging setup consisting of two best-form lenses (focal length $f$=15cm, diameter 25.4mm), resulting in the intensity distribution shown in Figure 1b. The resulting transient WGP is probed by a time-delayed THz pulse that travels collinearly with the pump beam after reflection off a pump-transparent glass plate with an indium tin oxide coating. The THz electric field is subsequently detected by electrooptic sampling, and the various polarization configurations (∥ and ⊥) are covered by appropriately rotating the shadow mask and the THz polarizers before and after the Si slab.

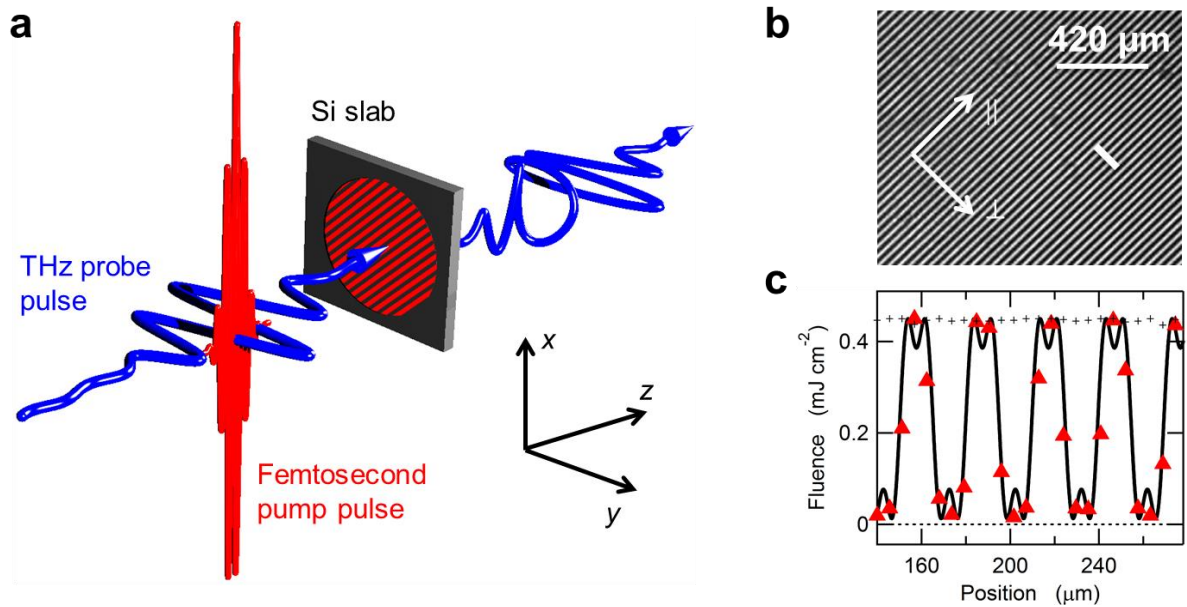

**Figure 1 | Scheme of a transient metamaterial induced by femtosecond gray-scale lithography. a**, A femtosecond laser pulse with a tailored beam cross section generates an electron-hole plasma with spatially varying density in a Si slab, thereby inducing a refractive-index modulation ranging from metallic to dielectric. The transient optical properties are probed by a time-delayed THz pulse. **b**, Intensity distribution of the laser beam cross section at the slab plane as obtained by imaging a striped shadow mask. **c**, Red symbols are the result of a scan along the white line in **b**, and the solid line is a calculation of the shadow-mask image based on the point spread function of the imaging optics used. The periodic pattern is expected to instantaneously induce a transient WGP that can be used to manipulate the polarization of a THz transient on a sub-cycle scale (see **a**). Part of the unshaped (homogeneous) beam profile is also shown (+).

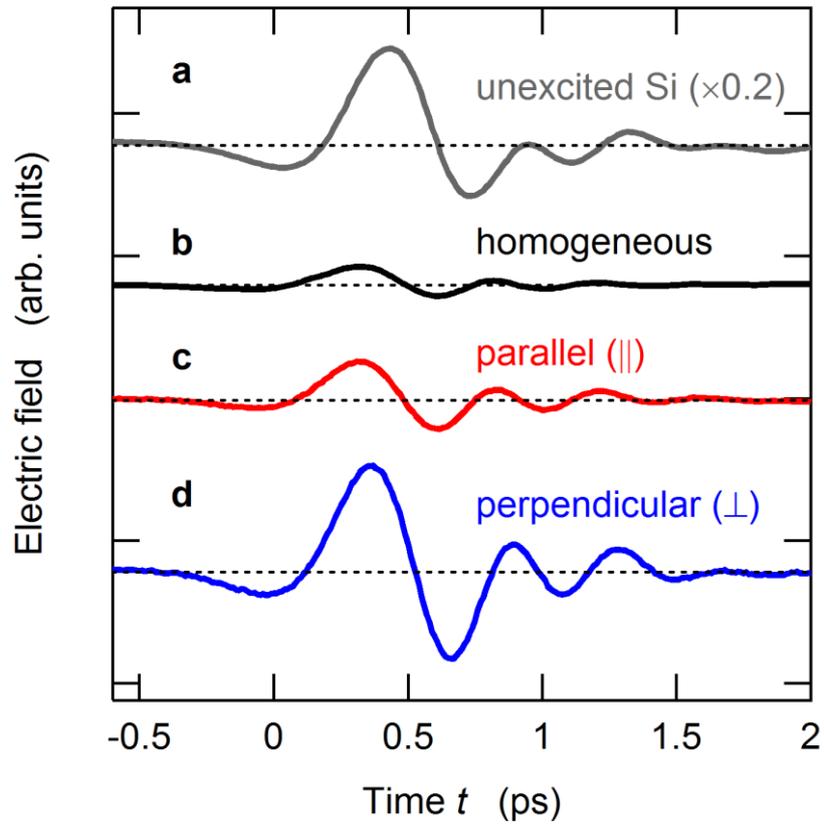

**Figure 2 | Time-domain characterization of the transient wire-grid polarizer. a**, Transient electric field of a THz pulse having traversed an Si slab without excitation **b**, 60ps after homogeneous excitation with an 800nm pump pulse, **c**, 60ps after excitation with the pattern of Fig. 1b with the THz polarization parallel and **d**, perpendicular to the wire orientation. The different amplitude and phase shift of the waveforms in **c** and **d** are a clear signature of the highly anisotropic response of the transient metamaterial.

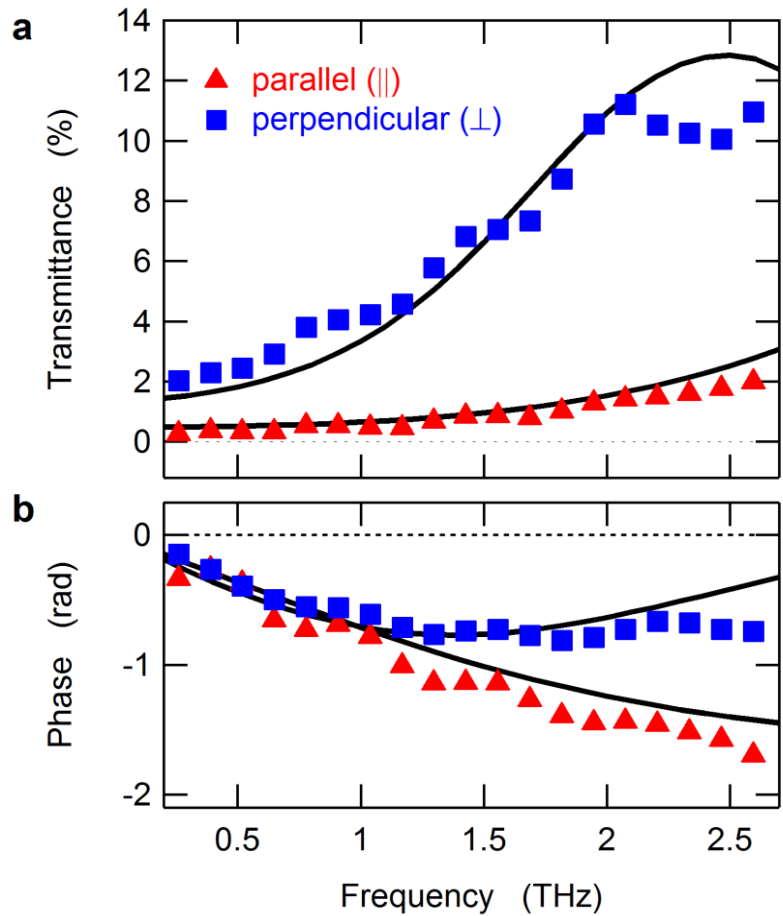

**Figure 3 | Spectral transmittance of the transient wire-grid polarizer. a**, Power transmittance of the Si slab 60ps after excitation with the pump intensity pattern of Fig. 1b, referenced to the unexcited slab. The spectra demonstrate that the transient structure acts as a polarizer. **b**, Spectral phase shift, revealing the more metallic behavior (negative phase shift) of the structure for THz radiation polarized parallel to the transient wires. Solid lines are model calculations based on effective-medium theory, without using any fit parameters.

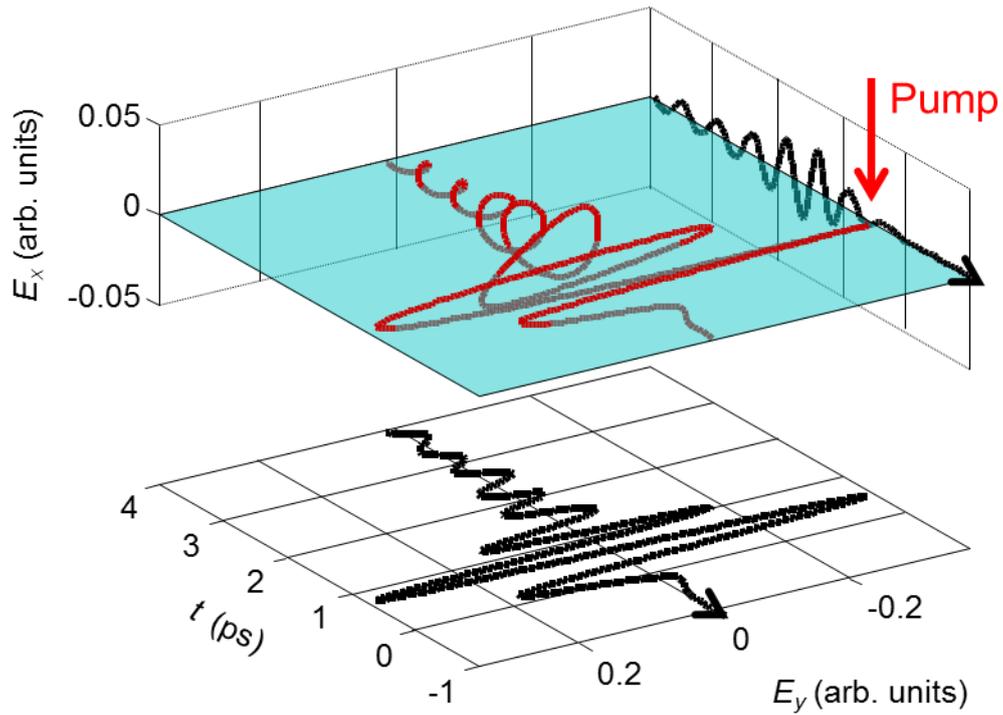

**Figure 4 | Experimental demonstration of on-the-fly polarization manipulation of a THz waveform.** The graph shows the measured transient electric field of a THz waveform after interaction with a time-dependent metamaterial (see Figure 1a). After the leading part of the THz pulse has traversed an unexcited Si slab, the transient WGP is switched on which permits changing the polarization of the trailing part from linear to elliptical. Units of $E_x$ and $E_y$ are the same.

# Supplementary Information

## Sub-cycle control of terahertz waveform polarization using all-optically induced transient metamaterials

N. Kamaraju, Andrea Rubano, Linke Jian, Surajit Saha, T. Venkatesan,
Jan Nötzold, R. Kramer Campen, Martin Wolf, Tobias Kampfrath

**Characterization of homogeneously photoexcited Si.** The refractive index of the Si slab (thickness 10µm) following homogeneous excitation with a femtosecond pump pulse (wavelength 800nm) is determined by a THz probe pulse (see Figure 1) as a function of the pump-probe delay. Since the THz dielectric properties change in a virtually step-like manner, it is sufficient to measure transmitted THz waveforms 60ps before and 60ps after photoexcitation (see Figure 2). After a Fourier transformation, we solve standard equations[1] of the transmission coefficient of a homogeneous slab for the desired refractive index $n(\omega)$. The resulting permittivity $n^2(\omega)$ is shown in Figure S1 for two different pump fluences. It can be described well by the so-called Drude model[2]

$$n^2(x, y, \omega) = \left(1 - \frac{\omega_p^2}{\omega^2 + i\omega/\tau}\right) n_\infty^2 \qquad (S1).$$

where $\omega/2\pi$ denotes frequency, $\tau$ is the mean collision time of the charge carriers, $\omega_p = (Ne^2 cZ_0/m^* n_\infty^2)^{1/2}$ is the screened plasma frequency determined by carrier density $N$, elementary charge $e$, light velocity $c$, free-space impedance $Z_0 \sim 377\Omega$, the effective mass $m^*$ of the Si conduction electrons (approximately 0.26 times the free-electron mass[3]), and the refractive index $n_\infty = 3.42$ of unexcited Si (ref. 4). Note that for plasma frequencies $\omega_p$ larger than the frequency $\omega$ of the probing THz radiation, $n$ attains a large imaginary part, implying strong attenuation and metallic behavior.

**Si reflectivity.** The refractive index obtained from transmission-type measurements (Figure S1) allows us to calculate the reflectance of the Si slab before and after photoexcitation. Results are shown in Figure S2.

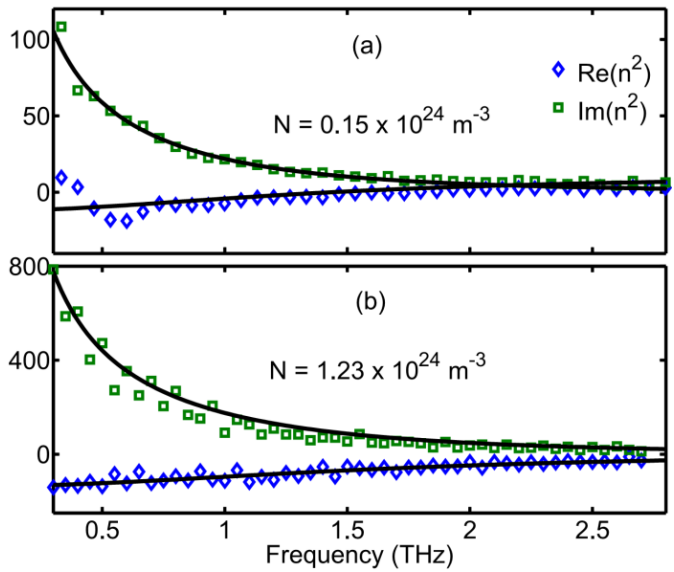

**Supplementary Figure S1 | THz permittivity $n^2$ of photoexcited Si. a**, At an incident pump fluence of 0.06mJ cm$^{-2}$, the negative Re $n^2$ indicates metallic behavior. A Drude fit [equation (S1)] yields a screened plasma frequency of $\omega_p/2\pi$=2.1THz and a carrier scattering time of $\tau$=115fs, which allows us to estimate a conduction-electron density of $N$=0.15×10$^{18}$cm$^{-3}$. **b**, Same as **a**, but for a pump fluence of 0.45mJ cm$^{-2}$, resulting in $\omega_p/2\pi$=5.8THz and the carrier scattering time $\tau$=95fs, implying a larger conduction-electron density of $N$=1.23×10$^{18}$ cm$^{-3}$.

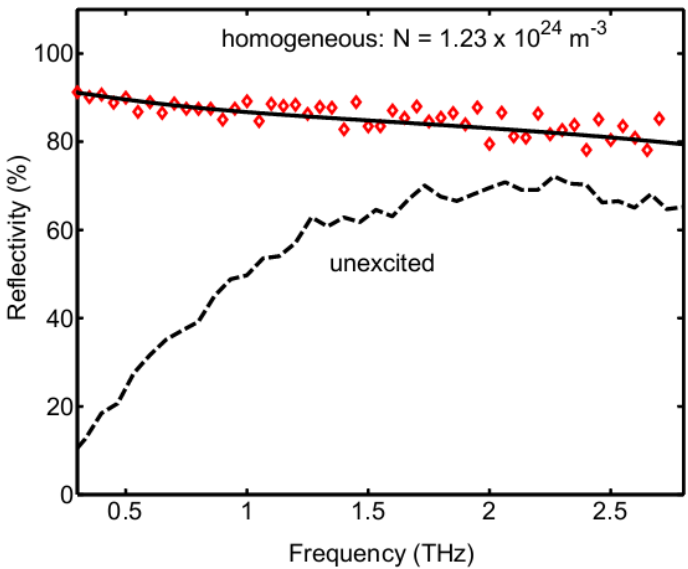

**Supplementary Figure S2 | THz reflectance of the unexcited and homogeneously excited Si slab.** Data (symbols and dashed line) are inferred from transmission-type measurements (Figure S1b), and the solid line is obtained for Drude parametrization of the Si refractive index (solid line in Figure S1b). The large reflectance found is a hallmark of metallic behavior in the frequency range covered.

**Calculated WGP performance under different conditions.** Using our WGP model (see main text), we calculate the WGP transmittance assuming a perfectly rectangular excitation profile (homogeneously excited and unexcited stripes) at temperatures of 300K (see Figure S1b) and 30K (ref. 5). Results are shown in Figure S3.

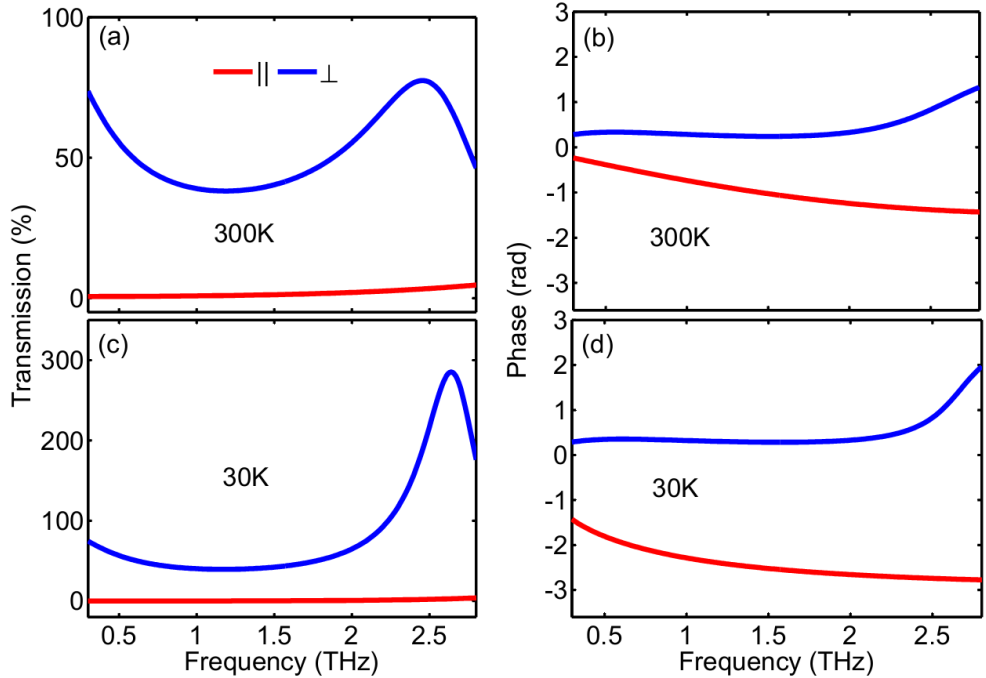

**Supplementary Figure S3 | Calculated WGP transmittance assuming a perfectly rectangular excitation profile. a,** Power transmittance and **b,** spectral phase shift at 300K, that is, with homogeneously excited stripes having the permittivity of Figure S1b and the other areas being completely unexcited. **c**, **d**, same as **a**, **b** but at a temperature of 30K. The plasma frequency of the excited rectangular stripes is $\omega_p/2\pi=5.8$THz as in **a**, **b**, but the Drude relaxation time of $\tau=2.1$ps (ref. 5) is more than one order magnitude longer than the $\tau=95$fs in **a**, **b**. The transmittance peak at ~2.5THz arises from a Fabry Perot resonance in the Si slab. Normalized transmittance values above 100% indicate that the transmittance (referenced to vacuum) of the excited slab is higher than that of the unexcited slab.